# ELECTROMIGRATION OCCURENCES AND ITS EFFECTS ON METALLIC SURFACES SUBMITTED TO HIGH ELECTROMAGNETIC FIELD: A NOVEL APPROACH TO BREAKDOWN IN ACCELERATORS.

## C. Z. Antoine[a], F. Peauger[a], F. Le Pimpec[b]

a) CEA, SACM,Centre d'Etudes de Saclay 91191 Gif-sur-Yvette Cedex, France
b) PSI, CH-5232 Villigen PSI, Switzerland

# 1. Summary

The application of a high electrical field on metallic surfaces leads to the well described phenomena of breakdown. In the classical scenario, explosive electron emission (EEE), breakdown (BD) originates from an emitting site (surface protrusion): the current at the apex vaporizes the emitting tip and the emitting current triggers a plasma in the vapor close to the surface. The plasma in turn melts the emitting site and makes it (hopefully) disappear. The conditioning process consists of "burning" the emitting sites one after another and numerous observations exhibit surfaces covered with molten craters that more or less overlap. In the case of radiofrequency (RF) applied fields, the effects of fatigue are also considered due to the cyclic nature of the applied stress. Nevertheless when dealing with RF cavities for accelerators, where higher fields are now sought, one can legitimately wonder if other physical phenomena should also be taken into account.

In particular, we believe that electromigration, especially at surfaces or grain boundaries cannot be neglected anymore at high field (i.e. 50-100 MV/m). Many publications in the domain of liquid metal emission sources show that very stable and strong emission sources, either ions or electrons, build up on metallic surfaces submitted to electrical fields through a mechanism that is slightly different from the usual localized breakdown evoked in accelerators. This mechanism involves the combination of electromigration and collective motion of surface atoms. In the case of emission source, this effect is sought after and has been extensively studied, whereas in our case it is very detrimental to the possibility of reaching high fields.

The recent results obtained on 30 GHz CLIC (Compact Linear Collider) accelerating structures, altogether with the data exposed hereafter have led us to propose a complementary scenario which could explain early melting of large areas of the surface.

In this paper we will concentrate on the early stages of breakdown, before plasma apparition. We will not discuss the plasma spot formation at the surface as we consider it to be the next step into the formation of the vacuum arc. We have gathered from the literature several examples of the physical phenomena involved on metallic surfaces submitted to very high fields. Definition of well-known concepts and terms used in other research fields will be introduced, like electrosprays, capillary waves…while some others have been left aside; not because they were irrelevant but because they would have requested extensive development which in turn would make this paper heavier. Because these concepts are, in a given community, well known a lot can be found using your favorite search engine and such without having to download the extensive bibliography cited in this paper.

In the introduction (section 2), we describe some of the damage that has been observed in CLIC accelerating structures which led us to suspect that electromigration is involved. We will present an alternative possible scenario for explosive breakdown (BD), which can result into the melting of extended area. In section 3 we will present RF simulations which show that pulse heating cannot be accountable for the observed melting. In section 4 we will describe what electromigration is and how it can lead to the appearance of nanotip and/or surface pre-melting. We will give several example of the occurrence of electromigration in several different experimental situations and we will try to evaluate some figures of merit. In particular we will show that electromigration is liable to occur at room temperature at fields close to 100 MV/m. We will also discuss other surface mechanisms that could also interfere with the breakdown mechanism.

A general discussion will be given in section 5 and the conclusion in section 6.

We hope to provide a new angle of observation that could help the accelerators community to better understand, and possibly overcome, the observed experimental limitation. Although it is very difficult to provide an evaluation of the relative weight of each phenomenon: electromigration, material type, surface state, plasma formation, we strongly think that the electromigration role needs to be explored.

# 2. Introduction

Several accelerating techniques are presently under development. Among the "classical" ones [1], the conditions chosen for the CLIC project are the most demanding in term of accelerating gradient, hence in surface field. Initially the project aimed at 150 MV/m accelerating gradient (270 MV/m peak surface field) at 30 GHz during 150 ns with a breakdown rate (BDR) lower than $10^{-7}$. After relative failure of the 30 GHz structures, specifications were reduced to 12 GHz with accelerating field about 100 MV/m (surface field ~ 200 MV/m during 200 ns pulses every 20 ms, with a breakdown rate (BDR) lower than $10^{-6}$ [1].

## 2.1. 30 GHz RF tests

RF structures with the so-called Hybrid Damped Structure (HDS) geometry were tested in several materials: Al, Ti, Mo and Cu [2, 3] (see Figure *1*). The HDS geometrical feature has the advantage of attenuating efficiently the high frequency signals induced by the electron beam with a relative simplicity of fabrication. Three accelerating structures (one for each material) have been fabricated in a quadrant version which requires only milling. No brazing is required since normally no RF current is present between quadrants [4]. As the structure is not leak tight, the full structure is placed inside a vacuum tank.

---

[1] By classical we mean RF acceleration, in contrast to plasma acceleration

High power tests have been performed in the CLIC Test Facility (CTF3) between September and December 2006 [5] with a 70 ns pulse length, for Ti, Mo and Cu structures. They respectively reached an accelerating gradient of 63, 51 and 61 MV/m (first cell) with a breakdown rate of $10^{-3}$. To reach a lower breakdown probability ($10^{-6}$), the field was decreased to 36, 42 and 42 MV/m [5] respectively. From Ref. [3, 5], we infer that with 70 ns pulses, the maximum surface field[2] was respectively ~ 92 MV/m for Mo and Al, 110 MV/m for Cu and 113 MV/m for Ti. Observation of the structures after RF testing has shown heavy damage at the surface, with large melted areas, whereas very high breakdown rate and heavy dark current have been measured during testing [2, 3]

Structures with different shapes (30HDS-TK.Cu [6] and 30CNSQ-TK.Cu [7]) have suffered high breakdown rates and observations after RF tests at respectively 110 MV/m and 135 MV/m surface field (SF) also shows some surface melting. This indicates that the problem needs to be addressed if higher fields with low BDR are required. On these structures, the field distribution is nearly the same. A comparison of these 3 different structures is shown in figure 1; the quadrants are meant to be in contact.

Structures have been changed from quadrant to disks to prevent the existence of gaps inside the high electrical field region. Indeed the breakdown rate at 30 GHz, 70 ns decreased from $10^{-3}$ to $10^{-5}$ ($10^{-7}$ being required).

All RF structures exhibit dark current after breakdown. Dark current is suspected to originate from field emission. In addition to beam perturbation, it can heat up the cavities' walls and trigger further breakdown, thence needs to be mitigated.

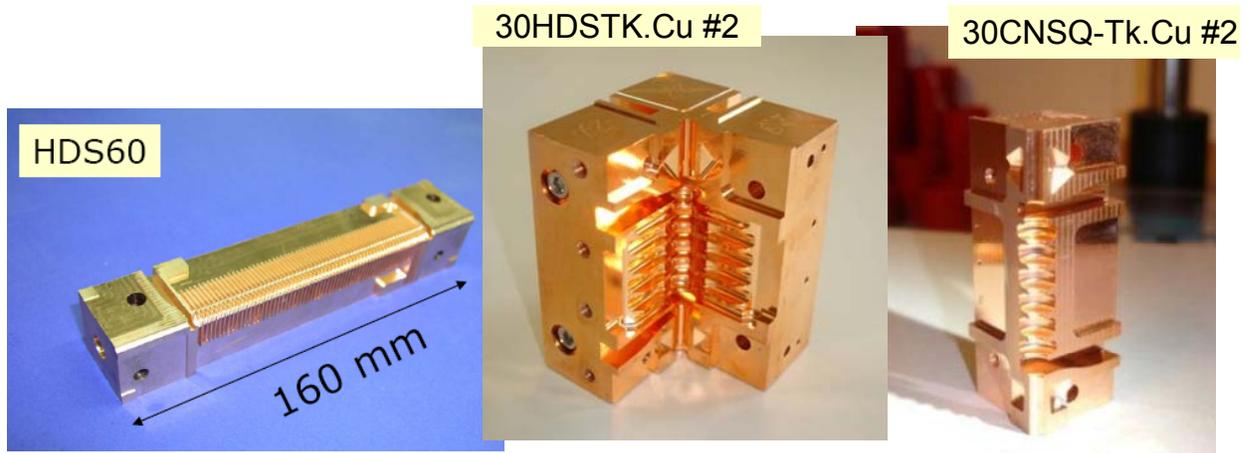

Figure 1 (color online): Pictures of the different structures evoked in the text. More details on HDS60 structures will be found in § 3

A change toward less stringent specification (180 MV/m peak surface field at 12 GHz, 240 ns pulses) was necessary, in regard of the technical difficulties to hold such high field (270 MV/m peak surface field at 30 GHz) with an acceptable breakdown rate [8].

The experimental facts, most of them published by CERN, which caught our attention, are presented in the following sections.

### 2.1.1. Large melted areas.

HDS60 structures made of Al, Ti, Mo and Cu were RF conditioned. Observation of the structures after conditioning, with surface fields between 95 and 135 MV/m, showed surfaces melted over a large area, especially on the $1^{st}$ cell where the electric field amplitude is maximal. As we will show hereafter, the RF power distribution combined with the thermal properties of the materials are insufficient to explain an apparent melting, at least if we consider the bulk melting temperature of the materials. (see §3). The melting appears to be too uniform to result from the overlap of numerous localized craters. Although the extended damage can be created by plasma spot during RF conditioning, we believe that one can consider other mechanisms, as exposed in § *2.1.3.*

---

[2] If not specified otherwise, we will always refer to surface field hereafter. We suppose $E_{surf}$ ~1.8 $E_{acc}$ for CLIC geometry.

### 2.1.2. Diffusion at grain boundaries?

In the particular case of Mo, so-called "cracks" have been observed, whereas on Cu, lumps or "hillocks" seem to protrude out from the surface (see figure 2). Such features require further exploration, and we cannot exclude at this point that they seem to be related to grain boundaries. Phenomena such as accelerated diffusion of atoms or vacancies at surfaces or interface (e.g. grain boundaries) are commonly observed in a wide range of situations and could be involved here. Another possibility is the premature evaporation of Mo oxides, $MoO_x$, which are known to start to evaporate around 400°C: as the grain boundaries tend to gather impurities, different phases can form preferentially along the grain interfaces.

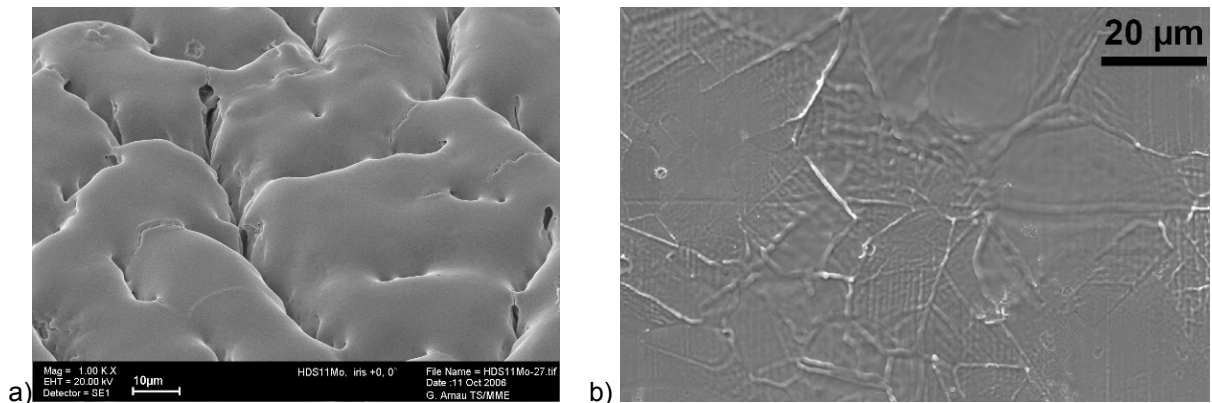

*Figure 2 : specific features observed on a) Mo [1] and b) Cu [7]. We believe that such features could possibly result from atomic diffusion processes at grain boundaries (enhanced diffusion of atoms and/or vacancies) resulting in voids or hillocks, depending of the material. Preferential diffusion and similar phenomena are currently observed at grain boundaries and surfaces over a wide range of conditions [9] (courtesy of CERN).*

Note that breakdown at grain boundaries was also observed at SLAC on similar Cu made RF structures. In this case segregation (diffusion) of sulfur compounds inside the grain boundaries was evidenced [10, 11].

### 2.1.3. Alternative scenario for large surface BD and melting

In the classical BD scenario (see Figure 3 and reference [12]), BD is initiated by a local defect (inclusions, dust particle, cutting flake…) which provide sufficient field enhancement to trigger field emission ($\beta$~10 to 300 [13]). The current at the apex of the emitting site becomes sufficient to heat up and vaporize the material. The emitted current can trigger a plasma by ionizing the vaporized material which generates the explosion of the emitting site, See Fig.3. The result is a local melting of the material with a typical diameter of a few µm. Large melted areas are believed to come from many overlapping craters. However, in reference [14] Wilson. claims that for a surface field close to 150 MV/m a single plasma spot could produce damages of 100 to 200 microns in diameter, when the heat is confined in the diffusion depth. Although we do not exclude such an explanation we believe that an alternative scenario could be considered to explain extended damage.

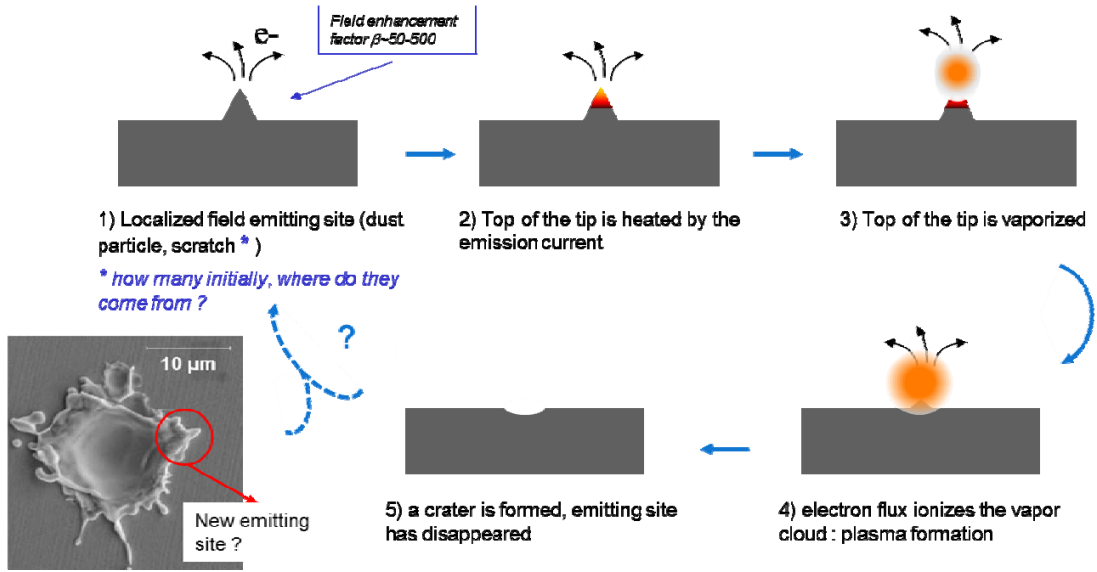

*Figure 3: Classical breakdown scenario: BD is initiated by a local defect (inclusions, dust particle, cutting flake…) which provide sufficient field enhancement to induce field emission. The current at the apex of the emitting site becomes sufficient to heat up and vaporize the material, where the emitting current can trigger a plasma and generate the explosion of the emitting site. This scenario does not consider the origin of emitting sites.*

The alternative scenario we propose here is based on the conjunction of electromigration and collective effects; the so called "capillary wave instability" Figure 4. This scenario was first evoked in the functioning of liquid metal emission sources and we will try to describe it in detail in section *4.2.3* and references cited therein.

In short, it involves the existence of small surface irregularities (some nm are sufficient) that give rise to a local electric field gradient. When the field is high enough, electromigration, as described § *4.1,* tends to enhance the surface irregularities because atoms tend to migrate along field gradient toward the top of the "hill" and form nanotips. In certain conditions e.g. in the presence of RF field in a certain frequency range, this effect becomes collective and helps to form an array of nanotips. Once this phenomena starts, it can only run away because the field gradient become stronger, which in turn enhances electromigration etc. until the surface becomes liquid like. The array of nanotips can soon become an array of field emitters, a strong source for dark current. If the field is further increased, explosive emission will appear not only on one isolated emitting site, but over a large number of tips altogether. Once the surface is molten it can re-form nanotips (Taylor cones) as long as the field is applied. Field emission, providing dark current, is then no longer suppressible. Dark current might in addition favor the apparition of BD since, by heating up the surfaces, it also enhances electromigration. With this scenario, the best way to prevent breakdown, as well as suppressing dark current and field emission, is to reduce electromigration, rather than using the classical conditioning sequences.

We believe that indeed the usual breakdown mechanism is insufficient to explain uniform, extended melting of the surface or the higher dark current observed after breakdown in normal conducting cavities; additional physical effects should be taken into account. We need thus exploring other type of mechanisms: either the local field is higher than the applied one, or the melting temperature is lowered by some mechanism. At this point, we must determine what are the actual field and temperature distributions, and find which other contributions might come into play.

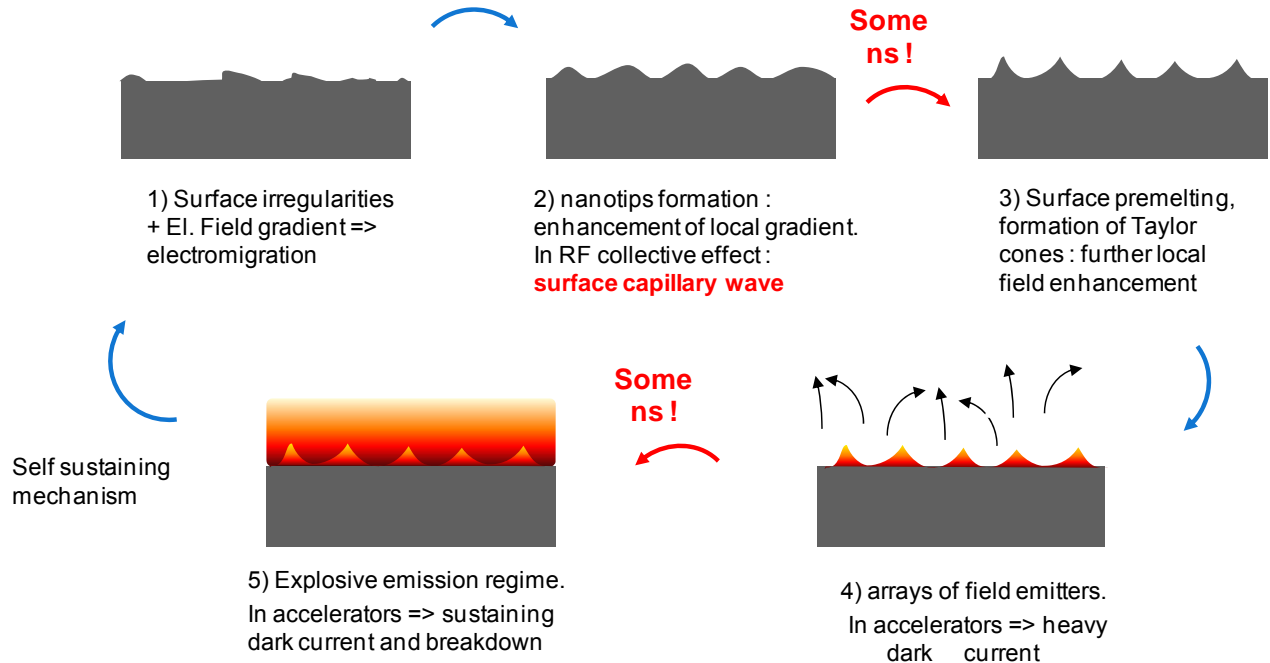

1) Surface irregularities + El. Field gradient => electromigration

**Some ns !**

2) nanotips formation : enhancement of local gradient. In RF collective effect : **surface capillary wave**

3) Surface premelting, formation of Taylor cones : further local field enhancement

Self sustaining mechanism

**Some ns !**

5) Explosive emission regime. In accelerators => sustaining dark current and breakdown

4) arrays of field emitters. In accelerators => heavy dark current

*Figure 4: Alternative scenario for breakdown: electromigration tends to enhance the surface irregularities because atoms tend to migrate along field gradient toward the top of the "hill" and form nanotips. In certain conditions, such as the presence of RF field, this effect becomes collective and will help form an array of nanotips. Once this phenomena starts, it can only run away because the electric field gradient become stronger, which in turn enhances electromigration etc. until the surface becomes liquid like. The electromigration scenario shows that in some conditions, emitters can be formed simply due to the presence of electric field and can trigger breakdown continuously (see text for details).*

# 3. RF simulations

As a first step it is important to demonstrate that the melting of the surface cannot be simply attributed to pulse heating.

The CLIC accelerating structures studied in this paper are travelling wave structures working at 30 GHz on the fundamental mode $TM_{010}$ with a phase advance per cell of 60°. Each structure has 11 regular cells and 2 special cells for matching with the input and output waveguides. The regular cells are identical so that the accelerating electric field is at maximum in the first cell and decreases exponentially along the structure due to ohmic losses. Each cell is coupled to four orthogonal waveguides in order to damp the higher order modes (see figure 3).

The structure has two symmetry planes. Perfect magnetic boundary conditions can be used on the xz and yz planes so that only one quarter of the structure needs to be simulated. The z axis is usually defined as the axis of the machine or the axis on which electron beam travels.

Finite conductivity (bulk) boundary conditions are put on all the metallic parts. The values for the different elements are: titanium ($\sigma=1.82\ 10^6$ S/m), molybdenum ($\sigma=17.6\ 10^6$ S/m) and copper ($\sigma=58\ 10^6$ S/m).

A single mode (the fundamental mode $TE_{10}$ in a rectangular waveguide) is considered at the input and output rectangular ports. This mode converts into $TM_{010}$ when propagating into a circular waveguide. The CLIC RF travelling wave structures can be considered as circular waveguides. A power of 20 MW at 29.9855 GHz is injected in the full structure which is approximately the power level achieved during the tests.

The mesh generator of HFSS [15] reproduces the curved faces with an imposed maximum deviation of 3 µm. Around $6.10^5$ elements are generated. A picture of the quarter of the structure and the structure after meshing is shown in figure 2.

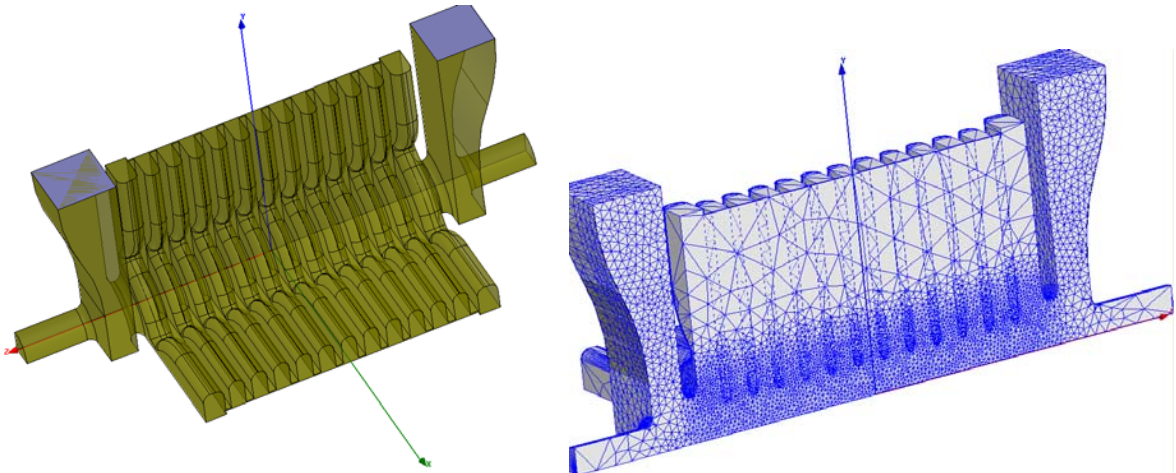

*Figure 5 : A quarter of the structure (vacuum) and the meshed structure*

For the three simulations (with Ti, Mo and Cu), the reflection coefficient between the input and output couplers stays below acceptable values: -28.6 dB for Ti, -38 dB for Mo and -42.6 dB for Cu (reflection in dB is $10\log P/P_0$, where $P_0$ is the incident power and P the reflected power. It is customary to consider this coefficient acceptable when bellow -25 -30 dB)

The transmission coefficients, which depend on the material conductivity, are -2.97 dB for Ti, -0.96dB for Mo and -0.53 dB for Cu.

The amplitude of the accelerating field is shown in figure 3. The electric field exhibits a small overshoot in the first and last cells (matching cells). This higher field in the first cell is due to the design of the accelerating structure. The accelerating field on axis in the first regular cell is about the same for the three structures and is around Ea=90 MV/m.

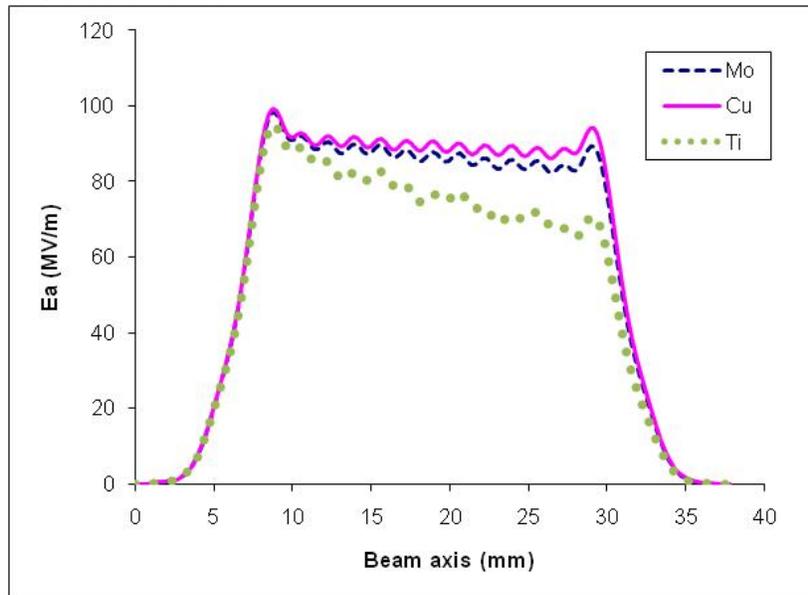

*Figure 6: Accelerating electric field on axis for Ti, Mo and Cu*

The surface fields and the corresponding temperature increase and current densities are given in table 1.

|  | Ti | Mo | Cu |
| --- | --- | --- | --- |
| Emax surf (MV/m) | 168.6 | 174.5 | 183.4 |
| Hmax surf (kA/m) | 333.6 | 341 | 356.2 |
| ΔT (°C) with Tp=70 ns | 853.9 | 18.9 | 23.7 |
| J (A/m²) | $\sim 10^{11}$ | $\sim 5.10^{11}$ | $\sim 10^{12}$ |

*Table 1 : Surface fields, temperature increase and current densities in the penetration depth*

The temperature increase is given by the following formula *(1)* [16]:

$$\Delta T = R_S \cdot \left| H_{surf} \right|^2 \cdot \sqrt{\frac{T_p}{\pi \cdot \rho \cdot C \cdot K}} \quad \textcolor{blue}{(1)}$$

where Rs, Tp, $\rho$, C and K are respectively the surface resistance ($\Omega$), the RF pulse width (second), the volumetric mass density of the material (kg.m$^{-3}$), the specific heat (J.kg$^{-1}$.K$^{-1}$) and the thermal conductivity (W.m$^{-1}$.K$^{-1}$). The increase of temperature is calculated for one single pulse. Since there is 200 ms between each pulse we can reasonably assume that the metal has time to cool down, at least for copper and molybdenum. Further calculation needs to be done for titanium to determine if its temperature gets higher with the accumulation of several pulses.

The current density is approximated by the ratio between the surface current density J$_{surf}$ calculated by HFSS [11] and the skin depth $\delta$ where most of the current is circulating:

$$J(A/m^2) = \frac{J_{surf}(A/m)}{\delta(m)} \quad \textcolor{blue}{(2)}$$

$$\delta = \sqrt{\frac{2\rho}{\varpi \mu}} \quad \textcolor{blue}{(3)}$$

where

> $\rho$ = resistivity of conductor
> $\omega$ = angular frequency of current = 2π × frequency
> $\mu$ = absolute magnetic permeability of conductor =$\mu_0.\mu_r$, where $\mu_0$ is the permeability of free space (4π×10$^{-7}$ N/A$^2$) and $\mu_r$ is the relative permeability of the conductor.

An example of the temperature distribution and the current density for Ti on the four first irises is shown in Figure 7. These distributions have been obtained using the formulas (1), (2) and (3) in the HFSS post processor. It shows that the surface reaches 1130 K in the highest current density zone, hence far from the melting point of Ti which is at 1941 K. Moreover, melted zones are rather observed in the high electrical field area, where the pulse heating stays moderate.

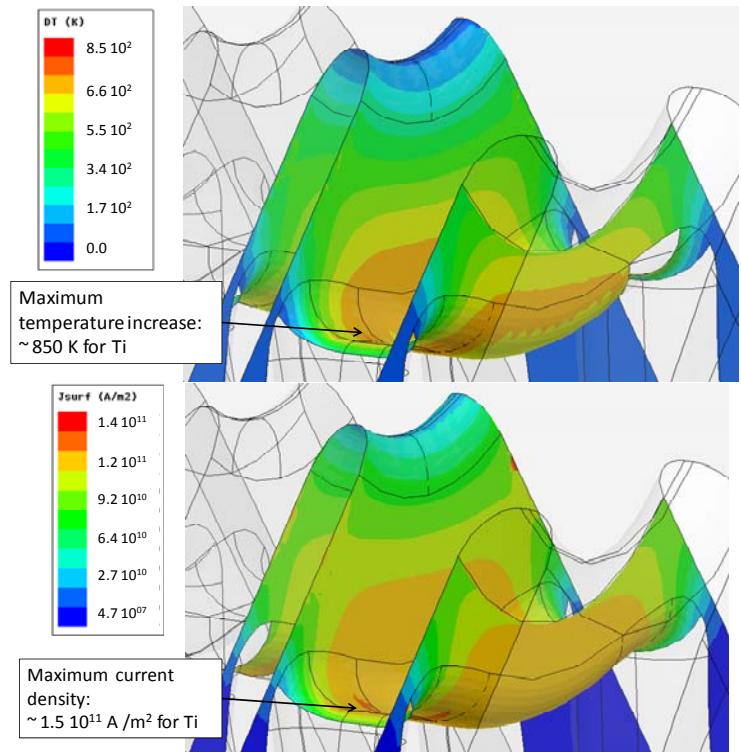

*Figure 7: (color online): temperature distribution and current density for Ti for an accelerating gradient of 90 MV/m (first cell) and a pulse length of 70 ns*

As we have seen in table 1 the combination of RF data and material properties such as thermal and electrical conductivities, show that huge differences can appear for the actual temperature reached at the surface, depending on the material in use. Nevertheless, the calculated temperatures for pulsed heating (~1130 K for Ti and ~320 K for Mo and Cu) are far lower than their melting temperatures (resp. 1941K for Ti, 2896 K for Mo, and 1358 K for Cu). Moreover, melting is observed close to the highest electric field area (lower magnetic field area) and not in the high current area (high magnetic field area). Hence RF heating of the surface cannot alone explain the apparent melting of the surface. Note that we have also made similar calculation on the new 12 GHz copper structure. The maximum current density and temperature gradient are similar: $J_{max} \sim 10^{12}$A/m² and $\Delta T_{max} \sim 40°C$.

# 4. Electromigration: Literature survey

As will be detailed below, electromigration (EM) is a particular case of atomic diffusion. More precisely one can distinguish diffusion which is a random process, where atoms can go in any direction and migration where the general direction of atomic movement is influenced by an external force. Transport processes in solid bodies include (1) chemical migration due to concentration gradients, (2) material migration caused by temperature gradients, (3) material migration caused by mechanical stress, and (4) material migration caused by an electrical field [17, 18]. This last case is often referred to as "electromigration" and is liable to appear whenever a metal is submitted to electric field (see § *4.1*)

Technically one should consider EM as being solely due to the effect of the electric field. However, in the presence of a high electrical field or high current density, EM should not be separated from the other transport phenomenon. High current density, high electrical field gradients or high thermal gradients are three conditions often encountered together inside RF cavities. We think that electromigration should be considered in the study of breakdown in accelerating RF structures.

Vacuum breakdown, or electrical breakdown, has been studied for nearly a century, and we are still unable to select electrically stable materials without specific testing for each application type.

Surprisingly they are only few other scientific domains available in the literature that deal with surfaces submitted to high electrical fields, and even less with radiofrequency fields. Up to now we have listed only seven that will be described in detail in the following paragraphs:

1. Field Ion Microscopy (FIM) and Atom Probe Tomography (APT) *where atoms or clusters of atoms are field evaporated at very high fields, but with very low current density and cryogenic temperature.*
2. Scanning Tunneling Microscopy (STM), *which is now used as a tool to displace atoms on surfaces. (Electromigration, high field, low current density, low (cryogenic) and room temperature.)*
3. Liquid metal emission sources (LMES/LMIS), *either electrons or ions. (high field, high current density, DC or RF, high temperature, but far below melting point).*
4. Electromigration in electronic connectors (*DC, low field, high current density, medium temperature)*
5. Field sintering *(a densification process used in powder metallurgy, with DC or RF, low field, high current density, high temperature).*
6. Laser irradiation at high flux.
7. Electrospray ionization *as used for example, in mass spectroscopy.*

We remind that we do not include phenomena involving plasma development in this survey, as we consider it to be a second step in the breakdown phenomena.

As will be detailed in § *4.2.3* and *4.2.5* , two techniques, LMES and field sintering, exhibit a RF behavior very different from the DC one. In particular Liquid metal sources (ions or electrons) have gone through a thorough theoretical survey. It is shown (see below for references) that RF favors the formation of "capillary instabilities" at surfaces, resulting in the appearance of a disorganized (melted) layer of a few atomic layers in thickness. Under the action of the electrical field, the surface tends to form an array of nanoprotusions that induce strong emission (electronic or ionic depending on the polarization). If the field is further increased, emission of droplets starts (a mechanism close from electrosprays formation-see *4.2.7*), which can further develop into plasma formation over large areas.

As a result, a localized breakdown mechanism is not the only possible explanation to the appearance of melted surfaces as well as Taylor cones, even on high melting temperature metals.

In the following, we have tried to list other phenomena capable of locally enhancing the field or changing the melting temperature, such as field enhancement on dust particles or surface pre-melting.

The references cited hereafter are only typical examples extracted from larger number of papers and do not pretend to depict thoroughly the presented subjects. They are meant to be a starting point for further bibliographic research, allowing the reader to become familiar with the key words from each domain.

| Technique | Field range | Current density range | Temperature range | Comments |
|---|---|---|---|---|
| CLIC (RF acceleration) | **RF 100MV/m** | $10^{11}$-$10^{12}$ A/m$^2$ | **RT to some 100°C** | Heating negligible except in the case of Ti |
| FIM/APT | DC 1-10 GV/m **100 MV/m** | ~ $10^{-4}$ A/m$^2$ | 20-100K **RT** | Electro-evaporation ~30 GV/m |
| STM | DC Typical :1 GV/m **Starts @ 100MV/m** | ~ $10^4$ A/m$^2$ | 4.5 to RT | **Migration of the less attached atoms (surface, interstitials)** |
| Vaccum microelectronics, LMES/LMIS | DC : 1 GV/m **RF : 100 MV/m** | $10^{11}$-$10^{12}$ A/m$^2$ | Starts at RT, heats up to several 100°C | **Surface melting observed @ 0.5-0.75 $T_m$** |
| Electromigration in electronic circuits | DC **<< 100 MV/m** 10-50 V/m ! | **$10^{10}$-$10^{11}$ A/m$^2$ Starts @ $10^8$ A/m$^2$** | 2-300°C | In RF once a void is created, it can only increase |
| Metallic powder sintering | **<< 100 MV/m** 10 V to MV ! | not measurable | Current is used to get surface melting | **RF faster than DC :** diffusion increases |

*Table 2 : Conditions where electromigration is suspected to happen. See bibliographic justification in the following chapters*

## 4.1. Electromigration, surfaces under electrical field

Electromigration (EM), (sometimes quoted electrotransport or electrodiffusion), is a diffusion phenomenon in a metal in the presence of an applied electric field [17-19]. It is involved in many physical phenomena and is well documented [17, 18, 20-23]. In summary, it has been discovered at the end of the 19[th] century. The initial interest concentrated on the fundamentals of self electromigration in pure metals as electromigration proved to be a perfect tool to probe the interaction between mobile defects and charge carriers (see below). The first application of electromigration was purification of interstitial elements (H, C, O, N…) in refractory metals during the 60's. Its scope of interest changed drastically in the late 60's when electromigration was identified as the source of several failure types in interconnecting lines in integrated circuits. The problem is still stringent with the continuous reduction of dimensions in microelectronic circuits. Although electromigration in circuits occurs in very specific conditions (thin films, high thermal conductive substrate, high temperature and high current densities…), much can be learned from the enormous set of data assembled in this community (see § _4.2.4_)

In the early 50's electromigration was studied across the phase diagram of specific alloys. It was shown that the atomic motion is not only determined by the electrostatic forces and that the direction of mass transport can be reversed by changing the alloy composition. It is in fact correlated with the main charge carrier (electrons or holes) present. Moreover the force experienced by a moving atom depends on the type of defects in the lattice and of the atomic configuration of the jumping paths, which renders the modeling of the system particularly complex [17, 24].

Electromigration is the result of the balance between two opposing forces, as described in equation 4, namely the "electric wind" and the "direct force". The electric wind is the exchange of momentum from one charge carrier to another. This includes the scattering of the conduction electrons by the metal atom and the momentum transfer of the moving metallic ion to other charge carriers. The direct force is the action of the electric field applied on the charge of the moving metallic ion.

$$F_{eff} = \mid e \mid . (Z^c + Z^w).E = \mid e \mid . (Z^*).E \qquad (4)$$

Where $e$ is the electron charge, $E$ the electrical field, and $Z^e$, $Z^w$ the respective contributions for each forces.

The balance between the electric wind and the direct force, i.e. effective charge Z*, depends strongly on the metal type, on its density of defects (crystalline structure, chemical purity), the nature of charge carriers (electrons, holes), the current density, the temperature range, the thermal configuration, etc.

As an example, Z*~ 1 in thin Al interconnects while it is ~100 for bulk Al [25]. Concerning the migration of a substitutional atom in a matrix of noble metals (Au, Ag, Cu), the electron wind force is dominant (5 to 10 times the direct force) whereas inside refractory and transition metals, Z* is small and can be either positive or negative depending on the external conditions [17]. As quoted in [26], the ballistic model which is often applied in modeling microelectronic failures fails to provide a quantitative description of the wind force for bulk EM on transition and noble metals and is only applicable within the free electron model.

Diffusivity is highly sensitive to temperature because of its exponential dependence. Loosely bonded atoms like those on the surface, or at grain boundaries, which exhibit a higher diffusion coefficient than the bulk ones, are expected to move preferentially in an electrical field.

High current density, but also high electric field or high thermal gradient can all contribute significantly to increase EM [17]. For instance in FIM or LMIS, field-induced surface-atom migration is currently observed whereas the current density is strictly zero [27].

### 4.1.1. Thermal aspects

Diffusion and its particular aspect, electromigration, are thermally activated processes; which means that the diffusion process follows a law in the form of:

$$D(T) = D_0 \, e^{-\frac{Ea}{RT}} = D_0 \, e^{-\frac{Qa}{kT}} \qquad (5)$$

Where the activation energy Ea is in J.mol⁻¹ and Qa in eV/atom.

Surface diffusion is at the origin of surface pre-melting (see below), of faceting and of grain boundary grooving when polycrystals are moderately heated. As can be intuitively grasped, the activation energy is expected to decrease from bulk self-diffusion to surface/interface diffusion then to surface electro-diffusion. Figure 8 shows an intuitive representation of the effect of the field on the activation energy.

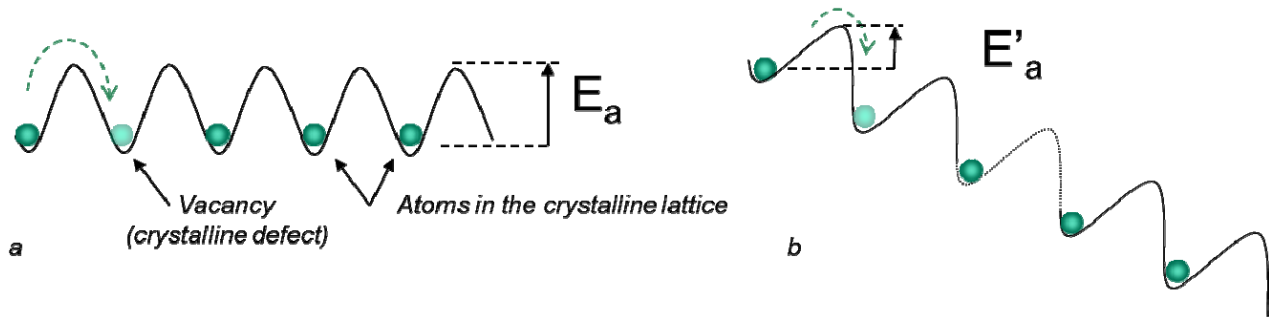

*Figure 8: Representation of the activation energy for atomic diffusion, and how it is affected by the presence of electric field. The presence of electric gradients leads to preferential diffusion.*

It is very difficult to get any actual figures since most of the diffusion data are measured at high temperature (typically ~ 75% $T_m$ (melting temperature)). Self-diffusion is expected to be negligible below room temperature. Nevertheless, Arrhenius plots hold over several hundreds of degrees for T < 0.75 Tm, and the activation energy (Ea) for atomic self-diffusion can be coarsely estimated with the following relationship [28] :

Bulk            Ea (J.mol⁻¹) ~ 146.5.Tm
Surface        Ea (J.mol⁻¹) ~ 55.4.Tm
where Tm is the melting temperature of the metal in K

In fact, it depends slightly on the crystalline structure and on the surface crystalline orientation of the material [29]. More accurate approximations can be found in [28, 30, 31]. Typical values of Ea (about 1 eV), independent of the crystalline orientation are summarized in table 2 after [28, 30-33] – We report on the order of magnitude of variations (when available) for temperatures ranging from room temperature to approximately 0.75 Tm, where the diffusion occurring by atomic jumps is the prevalent mechanism[3].

| Cu | Ag | Ti | Fe | Cr | Mo | W | V | Nb | Ta |
|---|---|---|---|---|---|---|---|---|---|
| 0.01-0.05 | 0.04-0.06 | 0.37-0.67 | 0.65 | 1.23-2.10 | 1.20-3.89 | 1.22-5.55 | 0.95 | 1.12-2.37 | 1.19-2.90 |

*Table 2 Activation Energy variation in eV*

Only a few figures are available on the comparison between field assisted surface migration compared to the situation without field. Most of the data comes from the FIM community (see § *4.2.1*) and concerns only high melting temperature metals. Usually the jumping probability is measured rather than the activation energy [34], and few atomic movements can be observed even at temperatures and fields as low as 78 K, 4 MV/m (Re onto W,). Reference [35] shows ab initio calculation of the activation energy for various diffusion mechanisms (hopping, substitution…) and shows that the hopping activation energy decreases with field (while it increases for other substitutional mechanisms). In reference [33], activation energy surface diffusion of tantalum is estimated to be 2.03 eV/At while it is lowered to 1.4 eV under electric field over 2000 MV/m. Reference [32] reports data for surface diffusion activation energies with and without (negative) bias around 0.5 MV/m. A decrease of 10-15 % of the activation energy seems to be systematically observed in presence of field. The same reference claims that the changes in the activation energy are expected to depend on the square of the applied field and that a greater field sensitivity is expected for high index surfaces (higher density of atomic steps and crystallographic defects) [32].

---

[3] The atomic jump mechanism is essentially a surface mechanism, involving an atom on the surface hopping from one crystalline cell to the adjacent. At higher temperature, inter-granular diffusion (at the grain boundaries) and then volume diffusion appear which then supplant the surface mechanism.

Typical electromigration activation energies as determined in electronic interconnectors, range between 0.5 to 0.8 eV for copper or aluminum (see § _4.2.4_ and [36-38]). These figures need to be compared with bulk EM activation energy : 2.3 eV for copper and 1.4 eV for aluminum [23]. In this case, temperature ranges are higher (300-500 K), measurement conditions are very different than in the FIM experiment and figures have to be taken with caution. In reference [39] it is found that the modification of activation energy is expected to be $10^{-8}$-$10^{-9}$ eV per mV.cm$^{-1}$ for surface electromigration on copper. If these figures were still valid for high fields about or above 100 MV/m, it would mean that the potential barrier would disappear at high fields and atoms would become free to move like in the liquid phase.

These activation energies are also strongly influenced by crystalline defects : screw dislocation emergence points are believed to reduce locally the activation energy by a factor 2 [40] and also by adsorption layers like organic contamination or oxides [41].

Every data tend to show that the activation energy can be influenced by several antagonist effects and that a better understanding could help to reduce the impact of the field on surface migration.

### 4.1.2. "Tip forming"

Changes in tip shapes have been thoroughly studied under the action of high electrical field and high temperature for various metals [40, 42-44]. Under the right conditions, faceting or crystal re-arrangement will occur and microprotrusions will form even on initially smooth surface [43-45]. Crystal re-arrangement happens when the balance between surface tension forces and the electric field forces is broken. Microprotrusions will form if the electric force overtakes the surface tension force, otherwise blunting of the tip will occur. The net surface migration current is independent of the electric field polarity [40, 43, 46, 47].

An analytical model, also applicable for broad-area cathodes, was developed by Bilbro [45]. The model considers a system with a finite tip radius and no lateral diffusion from the surface outside the considered radius. Bilbro shows that for intense enough electric fields, electrostatic energy will dominate surface energy and eventually the appearance of a nanoprotusion can minimize the system total energy. He shows that nanotips can form on initially smooth surfaces and that this model can satisfactorily reproduce the empirical power law relation between breakdown voltage and gap spacing usually observed in broad area electrodes ($V_{large\,gap}^{BD} \propto d^{m}$, 0.4<m<0.7 [45, 48]). In the case of nanotip formation, the local field is much higher than the external applied field, due to the curvature radius on the top of a defect of some ten nanometers. As we will see later on, in the case of surface melting, cooperative effects add up to this situation giving rise to a so-called "capillary wave", and several (many !) nanoprotusions can form at the same time.

Capillary waves arise at the interface between two fluids, a liquid and its gas phase for example, and are driven by surface tension, and gravity. This phenomenon was postulated in 1908 and described by Mandelstam in 1913 [49, 50]. The effects of capillary waves, in surface roughening, light reflectivity etc., have been thoroughly studied in liquid crystals, liquid metals as well as for water in a macroscale [51-53]. In the case of an RF structure, when near surface melting, thermal motion can produce such waves and at this scale (<1 mm) only the surface tension is relevant [54].

The possibility of re-shaping an arced tip is often used in order to extend the lifetime of a tip cathode [43, 55]. Blunting of tips can also be used to process single tip emitter or field emitter arrays (FEA). The goal of conditioning or processing is to obtain a FEA in which all tips are emitting homogeneously [56, 57]. Initially only a few tips of the FEA emit under the action of the applied electric field. Homogeneity in electron emission is achieved by firstly desorbing contaminants, adsorbed on the tip, by using the heat generated on the tip during field emission, or by adding an external source of heat, and by blunting the sharpest tips of the FEA, by gradually increasing the applied electric field.

Tip forming is usually controllable in the case of vacuum electronic devices, but is unfortunately not achievable in the present RF structures designs.

### 4.1.3. Surface pre-melting

Like every diffusion process, atoms with weaker bonding energies will diffuse faster, in particular atoms from the surface or from interfaces like grain boundaries. One of the main phenomena illustrating the specific behavior of surface atoms is surface pre-melting [58-61]. Indeed, unlike most of the phase transitions, melting does not exhibit a usual hysteresis behavior: it is possible to obtain under-cooling of a liquid whereas it is not possible to overheat the solid phase. This is due to the fact that most of the crystallographic faces start to soften and melt at temperatures around 0.75 Tm Surface pre-melting can also be considered like an order-disorder transition and originates of course from surface/interface diffusion. The thickness of the layer of concern depends on orientation, purity, surface curvature [59] and most of all, stresses either compressive or tensile. For instance it has been shown on Al (110) that the onset of surface melting occurs at 100 °C below the bulk melting temperature whereas it can be decreased a further hundred degrees under compressive stress [60]. In the case of RF cavities Laplace forces exert alternatively compressive and tensile forces that can certainly amplify any surface melting phenomena. Reference [59] discusses the relative influence of curvature, epitaxy, electric field, etc on the melting temperature of metallic surfaces. If nanotips form on the surface then their melting temperature is expected to be lower than the bulk one.

The effects of electrical fields will be more detailed in § *4.2*, but when an electrical field is applied to the surface, the chemical bonds of the first layer of atoms are distorted, favoring further their mobility.

### 4.1.4. Liquid surfaces under field

We have seen above that diffusion is greatly enhanced when the activation energy is low. In a general way, one can consider that the surface is "melted", in terms of ordering or mobility, when the diffusion coefficient reaches $10^{-5}$cm$^2$/s, even if it only concerns 1-2 atomic layers [62]. In this reference, Binh and Garcia put forward the idea of field surface melting at approximately one third of the bulk melting temperature,

Several other examples of field enhanced surface diffusion can be found without occurrence of vacuum breakdown (See § 4.2). Therefore we do suppose that in an operating RF structure **surface melting[4] can occur prior to heavy dark current emission and breakdown.** Liquid surfaces subjected to fields is a general problem thoroughly treated in the literature (see [63, 64] and references therein) and applies to a wide type of liquids including water [65], ionic and ferromagnetic fluids or molten metals. It has been particularly well studied in the case of a closely related problem: formation of electrosprays which are used for instance in mass spectrometers [66]. Conducting liquids subjected to fields tend to form cone-shaped structures (Taylor cones), stabilized by the balance of electrostatic and surface tensions forces [64, 67, 68]. If the field is increased the cone becomes unstable and the emission of micro-droplets starts [46, 47, 69-71], together with ions or electrons depending on the polarity [72]. In the case of molten metal the Taylor cones can be frozen by quickly switching off the voltage (see § *4.2.3*).

## *4.2. Electromigration occurrences*

### 4.2.1. Field Ion microscopy and Atom Probe Tomography (FIM/APT)

In these techniques, material samples are shaped into a sharp needle in order to locally enhance the electrical field to which they are submitted [73]. The experiments occur at low temperature (T~50-80 K) and very low current density. In FIM, adsorbed gas atoms are preferentially electro-desorbed from the less energetic sites of the surface and then ionized, while in APT [74] the surface atoms, including metallic cations from the tip itself are electro-evaporated. Typical applied fields are of the order of 30-50 V/nm, i.e. some 30-50 GV/m which is clearly much higher than CLIC conditions. Nevertheless, electromigration of surface atoms is observed at fields tenfold lower (note: the migration is assisted by the field gradient). Formation of nanotips by field-enhanced diffusion-growth on the apex of the FIM needles has been observed (e.g. [75-77]. In [78] Nagaoka et al claimed to observe it occurring at room temperature.

Indeed, increasing the temperature from 50 K to 300 K decreases the field for observed atomic migration by another factor 5 to 10 [79]. Since most of the FIM/ATP experiments are done with refractory

---
[4] « Melting » is used in absence of appropriate term. To our knowledge, "electrofusion" only applies to jazz music…

metals[5], the migration rate should be even more pronounced with lower melting point metals. Diffusivity is indeed highly sensitive to temperature because of its exponential dependence. Electromigration is thus likely to appear at fields around 100 MV/m at room temperature or slightly above. Note that at fields ranging from 50 to 500 MV/m, the force applied by the electric field is between 0.1 to 1GPa. This force is not enough to break the material, but it is enough to displace poorly bonded atoms, as was observed on liquid mercury [80]. Recently laser irradiation has been developed as a complement to ease the evaporation of atoms in APT. Formation of molten, multi protuberant surfaces has often been observed during the development of the technique [81].

### 4.2.2. Scanning Tunneling Microscope (STM)

Electromigration is also observed in STM at very low current density and typical fields of a few GV/m [82, 83], still very high compared to the applied field in CLIC structures. In some cases (creation of gaps [84], isolated atoms on surface [85]), displacements of atoms have been observed at fields as low as 100 MV/m [77, 85, 86]. The invoked mechanism is the change of hopping activation energy. As we have seen previously, the activation energy is greatly reduced by the polarization energy (in presence of electric field) and the diffusion can start at much lower temperature [77]. We remind the reader that below 0.75 Tm, diffusion occurs essentially through atomic jumps at the surface.

### 4.2.3. Liquid metal electrons and ions sources

Liquid Metal Electron and Ion Sources (LMES, LMIS) are used in a broad range of applications as diverse as flat low-voltage displays, surface analysis techniques like ionic microscopy, Auger, XPS, SIMS…, ion or electron sources for lithography (Field Ion Beam (FIB)) and even space propulsion [87]. One of the latest development domains is vacuum microelectronics, an alternative to classic solid state electronics, where the nature of the electrons motion is ballistic rather than hole and electron conduction inside a semiconductor. Since there is no dissipation of energy in the transport medium (vacuum), there is an ascertained advantage for high power, high frequency applications [72, 88]. LMES provide high brightness, monochromatic sources with very high intensity, in ions or electrons depending on the polarity. As will be detailed hereafter, the topic is well documented and theoretically explored. In addition, their RF behavior has been tested in conditions quite close to the situation with RF accelerating structures.

Liquid metal sources exhibit two regimes: a first one with moderate emission that switches suddenly to an "Explosive Emission" regime where the current gains several orders of magnitude and reaches eventually a saturation value. Once in this second regime, the emission is very stable and very bright. In this explosive regime, it has been shown that the surface is completely melted and forms so-called capillary wave instabilities [89, 90]. A multitude of Taylor cones bulge out the surface, each one forming an array of emitting sources with nanometer sizes. Electron and ion emission occurs by field emission and field evaporation mechanisms, and their close dispersion in energy shows that ionization occurs very close to the emitting surface, probably in the dense plasma that appears in the explosive regime. As mentioned in § *4.1,* the surface shape is stabilized by the balance of electrostatic and surface tensions and the formed array reaches a kind of steady state, that explains how very stable, high current emission level is reached. More details on the stabilization of the tips and ion emission processes can be found in [46, 47, 64, 67, 71, 88, 91-93].

The most surprising and important feature is that explosive emission can be excited on liquid metal sources in RF field at field gradients tenfold smaller than in pulsed DC field [89, 93-96]. The origin of this phenomenon is unclear, although the most frequently proposed mechanism is pre-heating of the surface through field emission. Indeed, it is now possible to build liquid metal sources with high melting materials like graphite (Tm = 3800 K) or Tungsten (Tm = 3695 K) [97-99], by exposing the high melting point needle to an electron flow from a filament. In the case of W, heating of the surface in the range 2000-2300 K is sufficient to provide the appearance of nm size emitters [99], i.e. 1000 to 1500 K lower than the bulk melting temperature. To reach this result, the tip is first exposed to very high field (explosive regime) where the plasma heats up the surface and the field is then decreased: an array of nanotips is "frozen" down and provides further emitting sites. This picture seems to be very close to what was observed on titanium CLIC cavities, as shown in Figure 9.

---

[5] High activation energy for migration

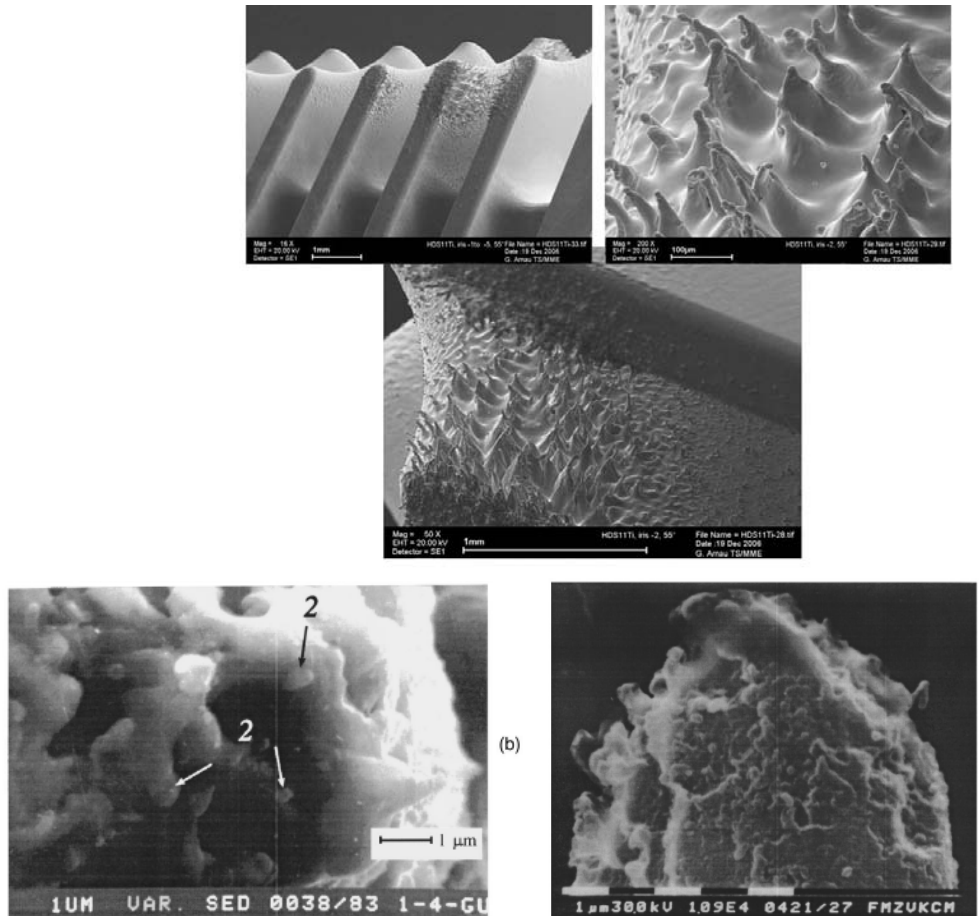

*Figure 9 : Comparison between surface state obtained on a titanium CLIC cavity (3 top pictures), and a copper (left down) and a graphite (right down) liquid emitter source (images from Cern (top) and [89] bottom).*

With the use of alternating electric fields, Popov [100] was able to create field emission electron sources operating **without the formation of explosive plasma** on various metallic surfaces such as stainless steel, tantalum, niobium or copper. He claims that compared to DC, the activation [onset of field emission] was ten times faster, the density of emitters was higher, and the current was more stable.[100].

There is also some other experimental evidences that the surface is already molten before the explosive regime is reached [101] and that nano-emitters persist after quenching (solidification) of the LMIS [102]

Although liquid emission sources often work in the explosive emission regime, the existence of plasma is perhaps not mandatory to obtain appreciable electromigration and formation of several field emitters.

### 4.2.4. Electromigration in electronic circuits.

DC electromigration in thin metallic connections is also well documented into the literature (see e.g. [17, 21-23, 103-108]). A few metals (namely aluminum, copper, and in a lesser extent gold) have been extensively studied. EM is known to be the origin of formation of voids in one side of the circuit and lumps (hillocks) on the other side. It is observed at a temperature of about 200 °C, very low voltages (some 10 to 100 V) [108], and with high current density, typically $10^{10}$-$10^{11}$ A/m$^2$. Indications of early electromigration have been observed at current densities as low as $10^8$ A/m$^2$ ($10^4$ A/cm$^2$), i.e. much lower than the current density applied to RF cavity surfaces. (As we have seen in § 3 the current density in CLIC structures ranges between $10^{11}$ A/m$^2$ and $10^{12}$ A/m$^2$ whereas it reaches several $10^{12}$ A/m$^2$ for superconducting Nb cavities).

The main mechanism in action in interconnects is momentum transfer from the conduction electrons (the electron wind term is dominant in these condition). Note that some papers from this community tend to reduce the term "electromigration" to electron wind and distinguish it from e.g. thermal gradient effects (see e.g. [109]).Reducing electromigration to this aspect is oversimplified [17, 23]. We have chosen to keep with the wider acceptance of the term.

EM is dominated by grain boundary diffusivity in aluminum, but for copper, EM results half from grain boundary and half from surface migration [110, 111]. The explanation relies on the fact that aluminum tends to protect itself with a strong oxide layer which blocks surface diffusion [23]. Little is known for other metals, but it is expected that grain boundary diffusivity will still be the main mechanism for temperatures lower than 0.5.Tm.

Surfaces and interfaces are known to strongly influence the formation of voids [104]. Pre-existing "voids", i.e. groups of vacancies, can often be found at triple points and are liable to ease the electromigration process [17]. This explains why these interfaces or non-uniform grain sizes usually promote EM and formation of voids[22]. Once a void is created, it can only grow further due to the different jump probability between vacancies and atoms[6]. The current density nearby increases since current flow needs to go around the void and the phenomena continues to amplify. Elastic and plastic deformation, stress, are also known to interfere [23, 105], probably due to their influence on migration rates.

EM also creates failures in AC operation [112]. Thermal gradient seem to play a paramount role in this case because Sorêt and Thomson effects[7] enhance migration rates [109, 113]. For AC, the momentum transfer is liable to be less efficient since the current changes its direction each half period, nevertheless holes were found to appear on strips at rates which are 0.1 to 10% of the DC rates [25], The formation of voids results from natural asymmetry: first of all, atoms and vacancies migrate along specific crystallographic directions, not necessarily parallel to the electron (holes) direction. In AC/RF the movement of atoms is "zigzag" like, which is why voids can only grow. Another source of asymmetry arises from surface crystallographic defects e.g. surface atomic steps (local change of diffusion) in activation energy [114].

In RF (12 GHz or above) the sign of the electric field switches in the time scale of sub ns which might be smaller than the migration time of the charge carrier. Nevertheless EM is still expected during a part of the RF cycle, under the action of the temperature gradient, the field gradient and the high current density.

Moreover, in the case of CLIC structures, the group velocity and the Poynting vector of the traveling wave, is known to play an important role [115]: momentum transfer is liable to have a preferential direction [116]. Electromigration could be the process at the origin of the cracks observed on Molybdenum on Figure 2a).

Among the recommended cures efficient in interconnects, we find:
- reduction of the current density,
- reduction of roughness and surface steps (construction as well as crystallographic) to prevent current concentration areas,
- reduction of grain boundary quantity (by enlarging the grain size for example),
- alloying,and
- encapsulating with a diffusion barrier on the top of the conductor.

Several of these routes could be explored in the context RF cavities.

## 4.2.5. Field sintering

Field sintering is used in powder metallurgy for the forming of high melting temperature material. It has many variants with various names (flash sintering, microwave sintering…etc), depending on specific technical details of the field/current application [117-121]. The field is used to internally heat up the powder and help its densification. In principle the heat repartition is more homogenous than with external furnace heating. The field can be DC, pulsed or RF… Since the system is very complex, little theoretical work has been done, although the acknowledged mechanisms seem to be mainly electrodiffusion (or electromigration) at the individual grain surfaces and is governed by power-law creeping. The driving sources for these material transport mechanisms are externally applied load, surface tension (sintering stress), and steady-state electromigration (electric field contribution to diffusion) [76].

---

[6] Atoms migrate preferentially along specific crystalline directions characteristic to the crystal structure of the solid. Even if one atom gets exactly the same momentum transfer in the reverse direction, it would not mandatorily move back along its initial trajectory, leaving a vacancy at its original place. Vacancies tend to gather, e.g. at grain boundaries.

[7] The Soret effect refers to preferential chemical diffusion in a temperature gradient, The Thomson effect is related to thermoelectricity.

It is nevertheless interesting to note that the microwave behavior is very different from the DC or even pulsed situation. In the case of RF, smaller applied electric field, lower temperature and shorter times are required to get efficient sintering, suggesting that some additional mechanism adds to the classical induction heating mechanism. The exact heating mechanism is still not fully understood considering the high reflective power of metals to RF and the very small skin depth compared to ceramics. Among the proposed mechanisms, eddy current losses in magnetic field, development of local hot spots, influence of dielectric losses on surface oxide layers and evolution of the RF absorption with temperature are evoked [117, 122].

### 4.2.6. Laser irradiation

For the case of laser irradiation, the surface of materials is also submitted to high intensity electromagnetic fields. Breakdown mechanisms have also been observed [123], and the formation of multiple conical structures have also been observed on many different materials (metallic or insulating) [124-126]. One interesting aspect is that the orientation of the melted cones depends on the orientation of the incident laser beam and lies parallel to the Poynting vector of the EM wave. In the case of 30 GHz structures, the molten area seems to appear in the area of maximum Poynting vector; a modified Poynting vector was also proposed in [116] to predict breakdown limitation.

### 4.2.7. Electrospray ionization

Electrospray ionization is widely used in analytical chemistry, e.g. in mass spectroscopy to get a uniform aerosol of very small droplets of the molecules to be analyzed. Dilute solution of analyte is pumped through a capillary. A high voltage (2–5 kV) is applied to the capillary. This voltage provides the electric field gradient required to produce charge separation at the surface of the liquid. As a result, the liquid protrudes from the capillary tip as a ''Taylor cone''. When the solution that comprises the Taylor cone reaches the Rayleigh limit, droplets that contain an excess of positive or negative charge are emitted from its tip [127].

## *4.3. Role of surface defects*

Most of the metals used for RF structure fabrication are covered with a natural oxide that can be either insulating or semiconductor (medium to large bandgaps). Natural metallic oxide exhibits a high density of defects either intrinsic, due to their structure (e.g. if they have a certain tendency to get amorphous), or extrinsic (impurities, dislocations, grain boundaries…). A mechanically polished or machined surface will even exhibit a higher density of foreign/displaced atoms, a higher density of dislocation, and of dangling chemical bonds.

Letting a fresh mechanically damaged surface, hence chemically reactive surface, in air will also increase defects: allowing irregular oxide growth due to the combination of strain and (humid) air oxidation. The repartition of these defects is not uniform. Such a growth mechanism helps to build a lot of localized defects into the bandgap, high carrier densities and some modifications of the Fermi level. These surface defects can locally modify the band structure as sketched in Figure 10 (right hand side), which leads to an apparently locally decreased work function.

The presence of traps in the forbidden band induces a curvature of the band at the interfaces as shown on the right hand of Figure 10. Since in most cases the Fermi level in insulators lies below that of metals, it was chosen to design it downward. Electrons can be injected from the metal and get accelerated by the local electrical field. Some of these electrons can even be directly emitted into vacuum, because they have gained an energy higher than the work function. In this case the mechanism of emission is thermionic and is expected to exhibit a higher emission rate at higher temperature. In support of this model is the observation that the emitted current in an RF cavity increases by about one decade between 77 K and 300 K [128], a fact not predicted by the classical Fowler-Nordheim model in this temperature range [129].

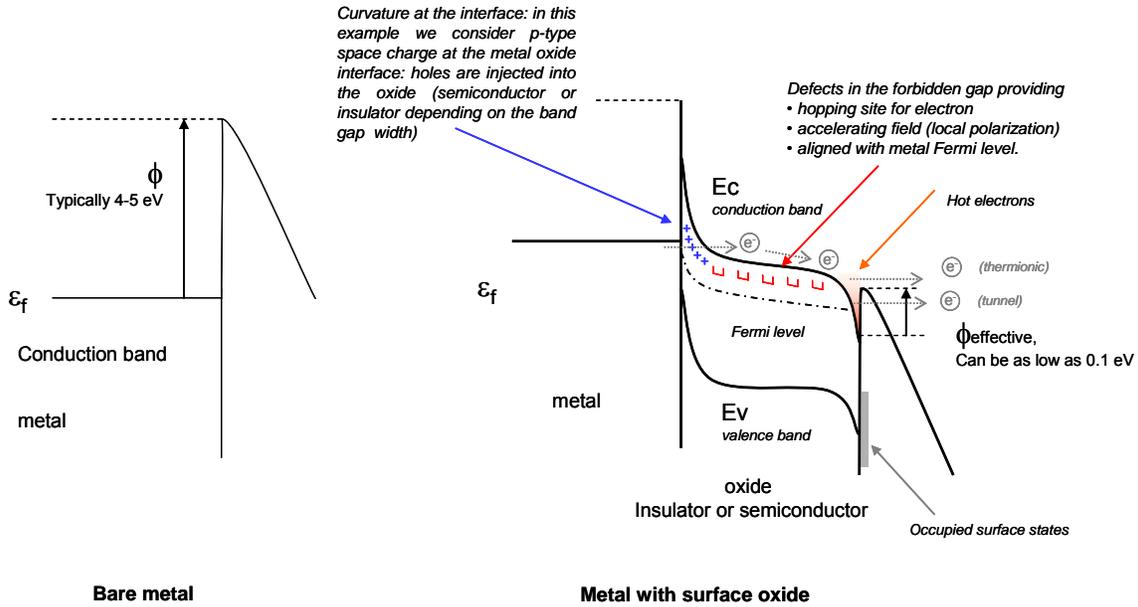

Curvature at the interface: in this example we consider p-type space charge at the metal oxide interface: holes are injected into the oxide (semiconductor or insulator depending on the band gap width)

φ
Typically 4-5 eV

$\varepsilon_f$

Conduction band

metal

Defects in the forbidden gap providing
• hopping site for electron
• accelerating field (local polarization)
• aligned with metal Fermi level.

Ec
conduction band

Hot electrons

$\varepsilon_f$

metal

Fermi level

Ev
valence band

oxide
Insulator or semiconductor

(thermionic)

(tunnel)

φeffective,
Can be as low as 0.1 eV

Occupied surface states

**Bare metal**                    **Metal with surface oxide**

*Figure 10 : Comparison of the near surface band diagram in the simplified model of a bare metal, and in the case of an insulating or semiconducting layer. The bending at the metal-oxide interface is due to adjustment between the Fermi levels from metal and oxide. Since in most cases the Fermi level in insulators lies below metals ones it was chosen to design it downward. The bending at the external surface is related to surface states due to e.g. adsorbed molecules. Here again the curvature depends on details of the crystalline orientation/polarizability of the oxide; we have sketched it in the most current situation. Defects in the gap can be related to foreign atoms, vacancies, crystalline defects like damage layer, etc… their density and energy distribution are highly dependent of surface preparation and are difficult to reproduce from one sample to another (After [130]).*

Chudnovskii et al [131] have studied the role of temperature in electroforming of Metal/Oxide/Metal (MIM) structure. Electroforming appears on several transition metals usually at lower temperature than breakdown: over a voltage $V_{th}$, a sharp and irreversible increase in conductivity is observed. It varies by about one decade from one position to another on the same sample, but there is a temperature $T_0$ at which $V_{th}$ tends to 0. This temperature is a characteristic of the metal which the oxide originates from. $T_0$ is comparable with the insulator–metal transition under electric field inside the metal sub-oxides. Sub-oxides are always present to some extent in high oxidation degree oxides. Defects like a dislocation or a dangling chemical bond not only increase the proportion of sub-oxides inside the oxide layer. Defects also increase the ionic and electronic conductivity of the oxide, and in particular the $O^{2-}$ mobility, helping the formation of "channels" with a modified conductivity.

This non-uniform behavior of oxide is often not enough to explain field emission by itself, it can nevertheless be the origin of the differences in results observed in emission and/or breakdown experiments done on samples prepared a similar way. Indeed one cannot master the surface states of samples prepared by mechanical polishing. Local "weak points" can appear due to the pile up of several local defects in oxide (stoichiometry variation, foreign atom, crystallographic defects…). In the case of mechanical polishing, such accumulation of defects is random and can vary a lot from place to place, even when the same preparation process is used. Reference [132] stresses the importance of patches of oxide in BD triggering.

Surface polishing is often used in mitigating dark current (field emission from material exposed to high fields [133]. Note that electropolishing and chemical polishing under proper conditions would leave a much more defined surface, especially in terms of crystalline damage. Recently at JLab [134] chemically etched monocrystalline niobium cathodes have been prepared to replace mechanically polished stainless steel (SS) ones. A smooth surface can be chemically achieved in less than one hour for monocrystalline structures (compare to several weeks for manual polish of SS). The niobium structures exhibit a 50% increase in achievable holding voltage compare to SS CEBAF DC cathode. For the purpose of a better basic understanding of the intrinsic properties of breakdown on metals, it would be desirable to test in situ prepared monocrystalline metal samples without any oxide, even if the preparation of these is hardly applicable to RF structure fabrication. A study on monocrystalline samples is currently ongoing within a collaboration between KEK, Helsinki University and CERN [135] as well as modeling of atom migration on surfaces under field by molecular dynamics [136].

Note that the advantage of surface polishing (mechanical machining), in term of dark current, is lost as soon as BD occurs as the surface smoothness is lost. After breakdown, we observe that electron emission threshold is lowered [137, 138]. We infer that local field enhancement can re-appear on the protuberances on the side of the craters.

# 5. Discussion

From what was presented above, it seems clear that electromigration and surface melting over large areas is possible (and observed) in the CLIC experimental conditions. In the classical breakdown scenario the melting occurs after arcing and is generally concentrated in a region of a few microns. When considering electromigration, there is a possibility that collective formation of several nanometer tips occurs prior to the breakdown through enhanced electromigration. Field emission is likely to promote some additional surface heating. Once the surface is molten, even on a couple atomic layers, many more surface emitters can appear. This in turn would explain the increase of dark current on cavities after breakdown and/or explosive emission. It is quite surprising that several papers point out a similarity between breakdown behavior in DC compared to RF (e.g. [139, 140]) since here there are numerous examples showing that RF worsens the situation, probably through pre-heating of the surface with field emission electrons impinging the surface.

The melted areas are usually not observed in the high current density area, but close to the high electrical field although not at its maximum, which suggests that current, field and temperature gradients all play a role. Breakdowns localized in a high magnetic field area, hence low electric field area, seem to be rather triggered by the presence of foreign material or perhaps by the presence of a gap between joint elements [[10, 141]].

In a recent and well documented paper, Jensen and co-workers tried to describe the thermal behavior of a multiple protrusion model [142]. They show that dark current observed in accelerators as well as in photocathodes can be attributed to field emission and enhanced diffusion. Unfortunately they didn't estimate precisely the origin of pre-heating of the surface, although they considered the Nottingham effect, resistive heating and electron/ion bombardment. Similar thermal calculations have been carried out for single emitters as well as for field array emitters in vacuum microelectronics [12, 88].

In superconducting cavities it has been shown that the power deposited by emitted area was around 1-2 W/cm$^2$ (see § _5.2_). In Ref [14] Wilson shows that electron bombardment, due to electrons coming from a plasma spot and accelerated back by the RF field to the surface, can produce substantial heating of the surface, but not enough to reach the melting point of the metal. His model could also apply to the initially emitted electrons produced by a field emitter. We think that a precise evaluation of the electron flux on surface just before breakdown would be very interesting.

Preparation in clean conditions, as it is done for superconducting cavities might also help to prepare emission free cavities. It would be interesting to determine if such cavities behave differently. Moreover, surface treatments like electropolishing should allow one to prepare more favorable, smoother, surfaces, less likely to retain contamination. On the other hand mechanical polishing hardens the surface and seems to help. The same effect is expected with alloying, provided that no eutectic mixtures with a lower melting temperature are formed.

Surface preparation is also often pointed out in breakdown studies and defects in the surface oxides (whether insulating or semiconducting) seem to play a paramount role. The existence of traps in the forbidden band is something to be expected for all oxide layers. Unfortunately their density depends a lot on the surface preparation and is fairly unpredictable. If such mechanisms are involved in the initiation of field emission, it is not surprising that there is such a wide spread in breakdown behavior of various materials, or even from one sample to another. For fundamental purposes, it would be interesting to be able to prepare flat surfaces without oxides and/or damage layers, e.g. through in-situ high temperature annealing (laser, electron beam…) before testing them for breakdown. As we mentioned, monocrystalline samples are recommended.

Capping surfaces with well determined layers could also help in gaining reproducibility. Indeed, recent pulsed experiment, 250 ns (FWHM) pulse length, on DLC (diamond like Carbon) coated metals (316L steel, Bronze or Cu) allowed one to reach, without breakdown, on steel 230 MV/m over a few mm gap compared to the max 100 MV/m on bare material [137, 143].

Recently, attempts have been made to cover a niobium RF cavity with a layer of alumina deposited with Atomic Layer Deposition [144]. The advantage of such layer is that it is very thin (some nm to some 10 nm) and amorphous, so there is nearly no strain in it. It acts like a perfect diffusion barrier and withstands high temperature. No field emission was observed on the cavity although it had not received any specific cleaning procedure.

Recently, Fursey and co-workers have discovered conditions where the apparition of explosive emission is impeded. It was observed on several metals as well as on carbon ([98] and references therein). It seems to involve the evaporation of all the nanotips. Once the surface is smoothened, up to the nanometer scale, it seems that external field alone is not able to promote surface melting and the formation of nanotips. Blunting of tips as described in 4.1.2 is also relevant to the same mechanism. We believe this effect provides a very interesting indication about the role of surface defects, and it shows that a regime where electromigration can be used rather than being detrimental may exist.

Finally, the best way to estimate the role of electromigration would be a systematic study of the initiation of breakdown as a function of temperature. To our knowledge, only very few experiment have been done in this specific context [145].

## 5.1. Relationship between breakdown rate and material type?

Breakdown rates increase with field until they reach an average steady value, which seems to be material dependent [140]. Many indications show that oxide layers are removed during the processing and that the steady state value is related to the naked metallic surface. One could easily imagine that the steady state value is related to one of the physical properties specific to each material. Indeed attempts have been made to use materials with higher melting point, in relation to their higher cohesion energy. Unfortunately this did not improve much the situation from the breakdown point of view.

Breakdown is obviously not related to just this aspect, at least not for DC (see e.g. [140, 146]). Many attempts to relate breakdown rates to material physical properties have failed. Even on the same material a large spread of results is observed, pointing out the difficulty of reproducible surface preparation. In addition, after the first breakdown, nanoprotusion can form on the surface with different field enhancement factors. Depending on their local shapes, it becomes very difficult to estimate the actual local field on the surface, which introduces another source of variation in the results.

Recently, Calatroni et al. showed, in a DC experiment on copper, that the quantity that is preserved after a few BD is the product $\beta.E_{BD}$ rather than the breakdown field $E_{BD}$ ($\beta$ is the field enhancement factor deduced from the Fowler-Nordheim slope of the field emission current measured after BD) [135, 147]. The authors, after a historical literature survey found that the relationship $\beta.E_{BD} = cons^t$ was already shown by Alpert et al fifty years ago [48], but seemingly forgotten. In addition, there is a correlation between the measured $\beta$ and the following $E_{BD}$: the higher the $\beta$, the lower the following BD field will be. They also noticed that when a surface is submitted to a constant voltage, BD tends to appear in clusters separated by "quiet periods". Measurement of $\beta$ slowly increases during a quiet period if the field is sufficiently high. When it reaches a certain threshold, BD systematically occurs. These recent experimental data are consistent with the existence of electromigration. The product $\beta.E_{BD}$ is probably related to the morphology a molten surface of a given material can produce in the presence of electromagnetic field. Several parameters might be involved, including the thermal properties of the metal but also various aspects like the surface tension of the molten surface, the influence of local adsorbed species etc. This may explain why, although the BD field is obviously something particular to a material, it is not easy to correlate it directly with its atomic physical properties only.

## 5.2. Breakdown, field emission and superconducting RF cavities.

Field emission being a very strong issue in superconducting cavities, it has been thoroughly studied in the SRF community (superconducting RF cavities for accelerators). Numerous experiments in DC and RF have shown a correlation between localized breakdown and field emitting particles [130, 146, 148-156]. In particular it has been shown that dust particles and scratches are good candidates for emission sites, and that the nature and the shape of the particles play a paramount role independently of the substrate material. In most cases, low level or no field emission is observed before breakdown but this suddenly increases during breakdown [138, 146, 150].

The presence of dust particles and roughness can locally increase the electrical field. Roughness due to machining or etching does not play much of a role: direct topological evaluation of beta in reference [157] and also measured with a tunnel microscope by Niederman [158] show that micron roughness induce betas of less than 10. Higher betas can only be attributed to the combination of defects at several size scales [142].

On the other hand it was clearly demonstrated that, due to their fractal nature, most of the natural dust particles exhibit nanometric defects that add up to the general shape factor, and field enhancement factor of about several 100 are easily achievable [146, 149, 159]. In [160], it was shown that when the current density reaches $10^{11}$ to $10^{12}$ A/m$^2$, the emitting area (in the Fowler-Nordheim equation) decreases to a few nm$^2$, a fact compatible with the formation of Taylor cone. Nevertheless dust particles are randomly distributed on the surface, and not all of them emit. If a dust particle can explain the limited melted zone like the few microns sized craters observed in the classical breakdown phenomena, it is difficult to figure out how they can trigger the melting of an extended zone several mm$^2$ wide.

In superconducting cavities, when breakdown occurs, it usually results in the formation of a single crater of a few microns, and it can either improve or decrease the cavity performance [148, 150]. Studies conducted on samples confirm this result, showing that processing is efficient only in 50% of the cases. In the other 50% of cases, welding of protruding particles onto the surface results in a very stable emitting site, hence a degradation of the cavity performance [130, 150]. With the improvement of cleaning techniques, surface fields up to 80-90 MV/m are currently observed without any breakdown or measurable field emission inside superconducting cavities [161]. The obvious difference between superconducting and normal conducting cavities is the operating temperature; a thermally activated phenomenon like electromigration being negligible at low temperature.

In addition, trajectories of field emitted electrons accelerated in various cavity types have been studied, and the resulting power deposition has been estimated. In reference [162] for instance, field emission was observed in a elliptical 1.3 GHz TESLA type cavity at ~40 MV/m surface field. The temperature mapping of the cavity during test shows that the temperature only increased by several degrees on the internal surface and that the power deposition was around 1-2 W/cm$^2$. Such power is obviously not enough to heat up surfaces exposed to emitted electrons to its melting point (p 103 of [163], [14]), but it would be interesting to evaluate the exact situation closer to breakdown.

# 6. Conclusion

We hope that this bibliographic review will help to highlight and clarify the complementary surface and material aspects involved in the initiation of breakdown phenomena. The following steps of arcing and plasma formation have been and are still thoroughly studied as described elsewhere (see e.g. [163] and[116]).We believe that electromigration plays a paramount role in the early stage of breakdown, in particular in the building up of nano-protrusions, and formation of arrays of field emitters.

There have been numerous papers on breakdown over the past decades, but only a few address materials and metallurgical aspects, which to our opinion cannot be neglected. We think the accelerator community could find great benefit in contacting fundamental metallurgists, material scientists or microelectronic circuit developers, who could help define better what the ultimate possible behavior of a given material is (a point obviously not known yet in spite of many efforts). There are strong indications that we are very close to this limit and that very high fields cannot be achieved in the present conditions. We believe that only additional fundamental work will give us an indication about the intrinsic RF breakdown limit for technical metals. This work would in turn help in defining new fabrication and conditioning processes better suited to the achievement of very high accelerating gradients in RF structures.

# 7. Acknowledgements

We wish to thank R. Forbes (University of Surrey), O. Groening (EMPA), G. Arnau Izquierdo and S. Calatroni (CERN) for helpful discussions. We are indebted to our CERN colleagues for providing original documents. We thank T. Garvey for proofreading this paper.